\documentstyle[12pt]{article}
\begin{document}

\begin{titlepage}

\vspace{3in}

\begin{center}
{\Large Boson Mapping and Nonlinear Response of Type-II Superconductors}
\vspace{.5in}

{\tt   H.D.Chen, W.P.Bai,  D.L.Yin$^a$, Y.L.Zhu$^b$, G.Li and C.Y.Li}\par

\vspace{2mm}

 {\it Department of Physics, Peking University, Beijing 100871}\par

\vspace{3mm}
$^a$~{\footnotesize E-mail address:~ydl@ibm320h.phy.pku.edu.cn}\par
$^b$~{\footnotesize E-mail address:~zhuyl@pku.edu.cn}

\end{center}
\vspace{1in}

\begin{footnotesize}
\begin{center}\begin{minipage}{5in}

\begin{center} ABSTRACT\end{center}

~~~~The vortices in a high-Tc superconductor with strong correlated
pinning centers have been studied
numerically using the mapping to charged bosons in two-dimensions(2D)
and the Monte-Carlo algorithm. Considering the viscous dissipation of moving
vortices we derived a nonlinear voltage response expression which describes 
different regimes and their crossover uniformly. This equation accords
with experimental results.

\end{minipage}\end{center}
\end{footnotesize}

\vspace{1in}
~~~PACS number:74.25  \par

~~~Keywords: {\footnotesize  High Tc superconductor, current-voltage
characteristics, flux pinning }

\vfil
\end{titlepage}
\eject
\rm
\baselineskip=0.35in
{\bf I. INTRODUCTION}\par
The static and dynamic properties of vortices in the mixed state of
high-Tc superconductors(HTSC) have been intensively studied both
experimently and theoretically in recent years$^{\cite{1}}$. For applying
superconductors in external magnetic fields, it is important to minimize
the dissipative loses of moving flux lines by improving the flux pinning
with some kinds of strong correlated disorder (material
inhomogeneities). The vortex dynsmics with strong
correlated pinning can be studied efficiently by exploiting
the mapping between vortices and 2D bosons$^{\cite{8}}$.\par

 Similar to the physics of flux lines in a pure system$^{\cite{2}}$,
the statistical mechanics of vortices interacting with columnar
pinning centers which are aligned parallel to the magnetic
field may be mapped into the quantum mechanics of charged bosons
in two-dimensions (2D)$^{\cite{2}}$.
Table I summarizes the analogy between the vortices system, with the tilt
modules $\tilde \varepsilon_1$ and thickness $L$(length of vortex),
and the corresponding 2D charged bosons system$^{\cite{3}}$.

\begin {footnotesize}
\begin {center}
Table I Boson analogy applied to vortex transport
\end {center}
\end {footnotesize}
\begin {flushleft}
\begin {tabular}{|c|c|c|c|c|c|c|c|}\hline
Charged bosons & Mass & $\hbar$ & $\hbar/T$ & Pair potential 
& Charge & Electric field & Current\\ \hline
Vortices & $\tilde{\varepsilon_1}$ & T & L 
& $2\varepsilon_0K_0(r/{\lambda})$ & $\phi_0$ & $\vec z\times \vec J/c$ & E(J) \\ \hline
\end {tabular}	
\end {flushleft}
\par

 In the Bose-Glass phase, the linear resistivity vanishes for low
external current $J\ll J_c$,  and the most important mechanics for
vortex transport is ``tunneling" between different columnar effect sites
via the deformation of a pair of ``superkinks"$^{\cite{4}}$.
This is very closely related to variable-range-hopping(VRH)
transport of charged carriers in disordered semiconductors$^{\cite{6}}$, 
and leads to the highly nonlinear expression.\par
   By further exploring the analogy to two-dimensions(2D) Boses
localized at randomly distributed defect sites, it leads to
a ``{\it coulomb}" gap in the distribution of the pinning
energies $g(\varepsilon)$ near the chemical potential $\mu$ which separates
the filled and empty energy levels$^{\cite{6}}$. In the limit of infinitely
long-range interaction, $\lambda \rightarrow \infty $, one would expect
$g(\varepsilon)$ to vanish near the chemical potential according to a power law
\begin {equation}
\label{eq.1}
g(\varepsilon)=|\varepsilon -\mu |^s.
\end{equation} 
This distribution affects the vortex transport properties. An inplane 
current $\vec J \perp \vec B$ induces a Lorentz force per unit length 
$\vec f_L$ perpendicular to $\vec J$, acting on all the flux lines:
\begin {equation}
\label{eq.2}
\vec f_L=\frac{\phi_0}{c}  \hat z  \times \vec J
\end {equation}
which modifies the free energy of vortices system.
In the boson picture this additional term represent an electric field
$\vec E= \hat z \times \vec J/c $ acting on the particles carrying charge
$\phi_0$. In the spirit of the thermally assisted flux-flow(TAFF) model of
vortex transport$^{\cite{12}}$, the superconducting resistivity
$\rho=E/J $  may be written as
\begin {equation}
\label{eq.3}
\rho \approx \rho_f \exp[-U_B(J)/kT]
\end {equation}	
where $\rho_f$ is a characteristic flux-flow resistivity, and
$U_B$ represents an effective barrier height which is of the
type $U_B(J)=U_0(J_0/J)^p$.$^{\cite{8}}$\par
Besides this inverse power-law $U_B(J)$, some other types have also
been suggested, such as the Anderson-Kim model
$U_B(J)=U_c(1-J_c/J)$$^{\cite{9}\cite{10}}$
and the logarithmic barrier $U_B(J)=U_c \ln (J_0/J)$ $^{\cite{11}}$.\par
In a system with moving vortices, the total current density 
$J=J_s+J_n$, where the normal component
\begin {equation}
\label{eq.4}
J_n \equiv E(J)/ \rho_f
\end {equation}
with $\rho_f \equiv \rho_n(B/B_{c2})$ which gives rise to dissipation and
viscous drag$^{\cite{13}}$. Thus the effect of Lorentz force on the effective 
barrier $U_B$(i.e., the saddel-point free-energy price for a flux
line to leave its columnar pin) should be attributed to the 
supercurrent component of the total current density
\begin {equation}
\label{eq.5}
J_s \equiv J-E(J)/ \rho_f.
\end {equation} \par
This paper is orgnized as follows. In the subsquent section we
briefly	describe the model and the Monte-Carlo simulation
procedure which have been succesfully used $^{\cite{3}}$. In section III 
we derived the current-voltage characteristics with the 
consideration of the viscous disspation of moving vortices. In section
IV we present results of simulation and compare them with some experimental results.
Finally, a short summary concludes this work.

{\bf II MODEL AND SIMULATION }\par
{\bf A.Model}\par
For describing a system which has $N_D$ columnar defect sites randomly
distributed on the $xy$ plane, we use the model in Ref.\cite{3} with  
two-dimensional effective Hamiltonian$^{\cite{3}}$ 
\begin {equation}
\label{eq.6}
H=\frac{1}{2} \sum\limits_{i\neq j}^{N_D} n_i n_j V(r_{ij})
+\sum\limits_{i=1}^{N_D}n_i t_i
\end {equation}
and its grand-canonical counterpart
\begin {equation}
\label{eq.7}
\tilde{H} \equiv H-\mu \sum\limits_{i=1}^{N_D}n_i.
\end {equation}
Here $i,j=1,2,...,N_D$ denote the defect sites. $n_i=0,1$ represents 
the corresponding site occupation number($n_i=1$ if a flux line is 
bound in columnar defect $n_i$), $\sum\limits_{i=1}^{N_D}{n_i}$ is the
number of the flux lines. $V(r)=2\varepsilon_0 K_0(r/\lambda)$ represents
the repulsive interaction between the lines; the modified Bessel function
$K_0(r/\lambda)$ describes a screened logarithmic interaction and
$\varepsilon_0=(\phi_0/4\pi\lambda)^2$ with $\lambda$ the London penetration
length. We have also included a random site energy $t_i$,
originating in the variation of pin diameters. 
Their distribution $P$ can be chosen to be centered at $<t>=0$,
with width $w$. For simplicity we assume a flat distribution
\begin {equation}
\label{eq.8}
P(t_i)=\theta(w-|t_i|)/2w
\end {equation}
[$\Theta(x)$  denotes the Heaviside step function].\par
For the interacting system(\ref{eq.6}), we define single particle 
site energies $\varepsilon_i $ as follows:
\begin {equation}
\label{eq.9}
\varepsilon_i \equiv \frac{\partial H}{\partial n_i}= 
\sum\limits_{j \neq i}^{N_D}{n_j V(r_{ij})} + t_i.
\end {equation}
For filled sites $(n_i=1)$, $\varepsilon_i$ is the energy required
to remove the particle at sites $i$ to infinity, for empty sites $(n_i=0)$,
correspondingly $\varepsilon_i$ is the energy needed to introduce an 
additional particle from infinity to sites $i$. In the thermal 
equilibrium, the chemical potential $\mu$ separates the occupied and 
empty states. With the intervortex repulsion taken into account, 
the distribution of pinning energies $g(\varepsilon)$ can be viewed 
as an interacting single-particle density of states and may be obtained 
from the statistics of the energy levels $\varepsilon_i$. 
One would expect that the normalization of the energy distribution is 
\begin {equation}
\label{eq.10}
\int\limits_{-\infty }^{+\infty }g(\varepsilon )d\varepsilon =N_D/A=1/d^2
\end {equation}
with $A$ the area of the system.\par

{\bf B.Simualtion}\par
Using a zero-temperature Monte-Carlo algorithm minimizing
the total energy with respect to all possible one-vortex
transfers$^{\cite{3}}$,
we reproduce the results of Ref.\cite{3} including the spatial configurations on
the ground states and the distribution of the pinning energy $g(\varepsilon)$.
We have performed extensive studies for the cases $N_D=200,400$. 
In order to study the size effect, we simulate with $N_D=800$ too. We 
have reproduced the result of Ref.\cite{3} of the size effect. Fig.1
shows our result of the energy distribution, where one finds the
``{\it Coulomb}'' gap.\par

{\bf III CURRENT-VOLTAGE CHARACTERISTICS }\par

Consider first the intermediate current regime $J_1<J<J_c$
($J_1 \equiv U_0/\phi_0 d$, and $U_0=<U_k>$ the average of pinning energy
$^{\cite{3}}$) in which
the motion of a single vortex is unaffected by the other vortex in the 
sample$^{\cite{4}}$. Driven by the external current $J$, a vortex will start
to leave its columnar pin by detaching a segment of length $Z$ into the
defect-free region, thereby forming a half-loop of transverse size $R$.
Considering the dissipation loses of moving flux lines, free-energy price of
forming a half-loop of transverse size $R$ by detaching a flux line
segment $Z$ into the defect-free region is apporixmately 
\begin {equation}
\label{eq.10.2}
\delta F(R,Z) \approx \tilde \varepsilon_1 R^2 /Z + U_0 Z-f_LRZ+
E(J)\phi_0RZ/c \rho_f
\end {equation}
From Eq.(\ref{eq.10.2}) we estimate the saddle-point free-energe
$\delta (F_1)^*$ and find the current-voltage relationship
\begin {equation}
\label{eq.11}
E(J)=\rho_f \exp [-(E_k/kT)(J_1/J_s)]
\end {equation}
where $E_k=d\sqrt{\tilde \varepsilon_1 U_0}$ and $J_s$ is the supercurrent
component described in Eq.(\ref{eq.5}).\par
For $J_L<J<J_1$($J_L$ is the current which satisfies $R^*(J_L)=L$),
we have to consider the configurational
limitation imposed by other vortex. The most important 
thermally activated excitation will now be a double 
superkink$^{\cite{3}}$. This is the vortex analog of variable-range-hopping 
charge transport in disordered semiconductor$^{\cite{6}}$. The cost in 
free energy for such a configuration of transverse size $R$ 
and extension $Z$ along the magnetic-field direction will 
consist of three terms:  (i)the double-superkink energy $2E_kR/d$
stemming from the elastic term, (ii)the difference in pinning 
energies of the highest-energy occupied site, $\varepsilon_i\approx\mu$ 
and the empty site at distance $R$ with $\varepsilon_i=\mu+\triangle(R)$,
and (iii) the viscous dissipation of the vortex stemming from the 
motion of the current around the vortex kernel in the external 
magnetic field $\vec B$, which is $E(J)\phi_0RZ/\rho_f$ $^{\cite{13}}$.
Thus the free-energy difference with respect to the situation without 
kinks and external current is
\begin {equation}
\label{eq.12}
\delta f \approx 2E_kR/d+Z\triangle(R)-f_LRZ+E(J)\phi_0RZ/c\rho_f
\end {equation}
The concentration available states as a function of $R$ with
$D$ dimensions transverse to $\vec B$ (here $D=2$) on the one hand 
equals $d^D\int\limits_{\mu }^{\mu+\triangle(R) }g(\varepsilon)d\varepsilon$,
and on the other hand is simply given by $\approx (d/R)^D$,
thus $\triangle(R)$ is to be determined from the equation$^{\cite{3}}$.
\begin {equation}
\label{eq.13}
\int\limits_{\mu}^{\mu+\triangle (R)}g(\varepsilon)d\varepsilon=R^{-D}.
\end {equation}
Optimizing first for vanishing current $J=0$ gives the longitudinal extent
$Z^*$ of the superkink as a function of its transverse size $R^*$,
\begin {equation}
\label{eq.14}
Z^* \approx - \frac{2E_k/d}{(\partial \triangle/\partial R)_{R^*}}.
\end {equation}
Upon balancing the last term in Eq.(\ref{eq.12}) against the optimized sum of 
the first two, one arrives at
\begin {equation}
\label{eq.15}
(J-E(J)/\rho_f)\phi_0/c=J_s\phi_0/c\approx\triangle(R^*)/R^* 
\end {equation}
which through inversion yields a typical hopping range $R^*(J)$.  
Inserting back into Eq.(\ref{eq.12}) finally yields the result for
the optimized free-energy barrier for jump,
\begin {equation}
\label{eq.16}
\delta F^*(J) \approx (2E_k/d)R^*(J)
\end {equation}
which we identify with the current-dependent activation
energy in Eq.(\ref{eq.3})
\begin {equation}
\label{eq.17}
E(J) \approx \rho_f J \exp [-(2E_k/kTd)R^*(J)].
\end {equation}
Considering a power-law form for the distribution of pinning energies
\begin {equation}
\label{eq.18}
g(\varepsilon)=\kappa |\varepsilon-\mu|^s
\end {equation}
one obtains
\begin {equation}
\label{eq.19}
\delta F^*(J)=2E_k(J_0/J_s)^p
\end {equation}
where the transport exponent $p$
\begin {equation}
\label{eq.19.2}
p=\frac{s+1}{D+s+1}
\end {equation}
and the current scale
\begin {equation}
\label{eq.20}
J_0 \approx c/\phi_0 \kappa^{1/(s+1)}d^{1/p}.
\end {equation}
Thus we get 
\begin {equation}
\label{eq.21}
E(J)=\rho_f J \exp [-(E_k/kT)(J_0/J_s)^{p}]
\end {equation}
with $J_s$ described by Eq.(\ref{eq.5}).

Eq.(\ref{eq.21}) can also be expressed in a general form$^{\cite{16}}$
\begin {equation}
\label{eq.22}
y=x\exp[-\gamma (1+y-x)^p]
\end {equation}
with 
\begin{eqnarray*}
\gamma & \equiv & 2p\ln \frac{J_L}{J_{Lf}}=2p \left (\frac{E_k}{kT}
\right )\left (\frac{J_0}{J_L-J_{Lf}}\right )^p\\ 
&  \approx & 2p \left (\frac{E_k}{kT}
\right )\left (\frac{J_0}{J_L}\right )^p\\ 
x & \equiv &\frac{1}{2} \left (\frac{E_k}{kT}
\right )^{-\frac{1}{p}}\left (\ln \frac{J_L}{J_{Lf}}\right )^{\frac{1}{p}}
\left (\frac{J}{J_0} \right )=\frac{1}{2}
\left (\frac{J}{J_L-J_{Lf}}\right )\\
&  \approx & \frac{J}{2J_L}\\ 
y & \equiv & \frac{1}{2} \left (\frac{E_k}{kT}
\right )^{-\frac{1}{p}} \left (\ln \frac{J_L}{J_{Lf}}\right )^{\frac{1}{p}}
 \left (\frac{E(J)}{J_0\,\rho_f} \right ) \\ 
& = & \frac{1}{2} \left ( \frac{E(J)}{(J_L-J_{Lf})\rho_f} 
\right ) \\
&  \approx & \frac{E(J)}{2J_L\rho_f}
\end{eqnarray*}
where $J_{Lf}\equiv E(J_L)/\rho_f$ which is much smaller than $J_L$.\par

Because of the size of the sample and the periodic boundry condition, 
the transverse size $R^*(J)$ cannot be greater than the sample width $L$.
For $J<J_L$, one would expect that the current-vlotage characteristics
will become Ohmic $^{\cite{4}}$.
This regime is also described by Eq.(\ref{eq.22}).\par
It is interesting to note, that the general form Eq.(\ref{eq.22}) is also
compatiable with other type of suggested $U_B(J)$. For instance, an $E(J)$
equation for type-II superconductors has been derived in connection to
the Anderson-Kim model with replacing the total transporting current density
$J$ in $U_B(J)=U_c(1-J_c/J)$ with $J_s$ of Eq.(\ref{eq.5})$^{\cite{14}}$, as
\begin {equation}
0\label{eq.Bai10}
E(J)=J\rho_f \exp[(-U_c-W_v+W_L)/kT]
\end {equation}
where $W_v=E(J)\cdot B\cdot {\cal A}/\rho_f$ is the viscous dissipation
term of flux motion, $W_L$ is the energy due to Lorentz driving force,
$W_L=J\cdot B\cdot {\cal A}$, the parameter ${\cal A}$ is a product of
the volume of moving flux bundles and the range of force action.\par
Eq.(\ref{eq.Bai10}) can be expressed with a reduced form
\begin {equation}
\label{eq.Bai11}
y=x\exp[-\gamma(1+y-x)]
\end {equation}
with $\gamma \equiv U_c/kT$, $x\equiv W_L/U_c$ and $y \equiv W_v/U_c$,
which corresponds with Eq.({\ref{eq.22}}) with $p=1$.

{\bf IV RESULTS  AND  DISCUSSION}\par

For simplicity, we assume that $E_k/k=1$. Thus temperature has unit $E_k/k$.
Moreover,in this paper the energy is normalized by $\varepsilon_0$, current
is normalized by $\phi_0d/\varepsilon_0c $ and voltage is normalized by
$\rho_f$. Using the relations (\ref{eq.13})-(\ref{eq.17}) and the
Monte-Carlo result $g(\varepsilon)$, we can numerically evaluate the
current-voltage characteristics for any form of $g(\varepsilon)$.
Results derived from the distribution of interacting pinning 
energies in Fig.1 are shown in Fig.2 which  has highly 
nonlinear characteristics and a superconducting phase. 
The double-logarithimic plots of $E(J)$ as the function of 
the current $J$ in the inset of  Fig.2 is consistent with 
the result in Ref.\cite{15}.\par
                                
From Eq.(\ref{eq.17}), we have
\begin {equation}
\label{eq.25}
\sigma \equiv \frac{d\ ln(E/J)}{d\ lnJ}=-(2E_k/kTd)\frac{dR^*(J)}{dJ}.
\end {equation}
This immediately leads to
\begin {equation}
\label{eq.26}
\frac{\sigma}{\sigma_{max}}={\frac{dR^*(J)}{dJ}}/
\left[\frac{dR^*(J)}{dJ}\right]_{max}.
\end {equation}\par
From Eq.(\ref{eq.15})-Eq.(\ref{eq.17}), we see that temperature will
have little effect near the maximal slope point.

On the other hand, we get from the analytical equation Eq.(\ref{eq.22}) 
\begin {equation}
\label{eq.22.1}
S\equiv \frac{d\\lny}{d\\lnx}=1+\sigma=\frac{1+p \gamma x(1+y-x)^{p-1}}
{1+p \gamma y(1+y-x)^{p-1}}.
\end {equation}
At the maximal slope point $(x_i,y_i)$, we have
\begin {equation}
\label{eq.23}
p^2 \gamma ^2 xy(1+y-x)^{2(p-1)}=1-p(x-y).
\end {equation}

As we have mentioned in the previous section that the Eq.(\ref{eq.22})
is compatible with different types of suggested $U_B(J)$, one would expect
its general agreement with experimental results of various kinds of type-II
superconductors.\par

In Fig.3 we compare the current dependency of the slope
$\sigma$ derived from Eq.(\ref{eq.22}) and Eq.(\ref{eq.23}) with the wide range
$V \sim I$ data observed by Repaci $ \it {et\ al.} $ on the YBCO films$^{\cite{17}}$.
The agreement is rather well.
\par
We also compared our Eq.(\ref{eq.22}) with the scaling of isothermal $E(J)$
curves observed by Koch $\it {et\ al.}$$^{\cite{18}}$. Taking simple trial form
$\gamma=10 (1-T/T_c)^\delta \cdot T/T_c$ with $\delta=0.5$ and $p=0.6$,
we get from Eq.(\ref{eq.22}) one hundred $E(J)$ isotherms near the 
$T_g\approx 0.84 T_c$(as observed in Ref.\cite{19}) with $T$ ranging from $0.74T_c$
to $0.94T_c$. All the isotherms collapsed nicely onto two curves($T>T_g$ and
$T<T_g$), consistent with the scaling of $\nu=1.7$, $z=4.8$ as shown in Fig.4.
The similar scaling result of Ref.\cite{18} is shown in the inset of Fig.4 which
has the same scaling exponents of $\nu=1.7$ and $z=4.8$.
\par

{\bf V SUMMARY}\par
Using the Monte-Carlo simulation method introduced by Ref.\cite{3},
we studied the current-voltage relation of type II superconductors
which have columnar defect sites. Considering the viscous 
dissipation of moving vortices, we find that the three regime's
current-voltage relation can be described by a unified equation
Eq.(\ref{eq.22}). This equation is consistent with different types of $U_B(J)$
so far suggested by different models and agrees with experimental observed 
$E(J)$ data of wide range of temperature and current density. With proper trial
effective barrier function $U_B(T)$ it gives $E(J)$ isotherms which can be
collapsed onto two curves ($T>T_g$ and $T<T_g$) with scaling exponents found
from experimental data.
\par

{\bf ACKNOWLEDGMENT}\par

	This work is supported by the Chinese NSF and the National Center
	 for R\&D on Superconductivity of China.

\newpage
\begin{center}
{\bf Figure Caption}
\end{center}
\begin{description}
\item [{\rm FIG.1.}] Normalized distribution of pinning energies 
$g(\varepsilon)$ as function of the single-particle energies
$\varepsilon$ which is normalized by $\varepsilon_0$ .
Another parameters is $w=0.2$. $g(\epsilon)$ vanished near 
$\mu$ according to a power law $|\varepsilon-\mu|^3$,
thus the power $p$ in Eq(19) is about $2/3$.

\item [{\rm FIG.2.}]  The nonlineal current-voltage curve
derived from FIG.1.$E(J)$. Voltage is normalized by $\rho_f$
while currrent is normalized by $\phi_0d/\varepsilon_0 c$.

\item [{\rm FIG.3.}] (a) The original experimental result of 
Ref.\cite{17}. 
(b)The result $\sigma/\sigma_{max} \sim ln(I/I_i)$
ploted from the results of Ref.\cite{17} are compared with
analytical Eq.(\ref{eq.22}) and Eq.(\ref{eq.23})(solid lines).
The parameter $p$ is taken as $1.8$ and $\gamma$ ranges from
$0.5$ to $5.0$ in $0.5$ intervals. 
$\sigma(I_i)\equiv\sigma_{max}$.

\item [{\rm FIG.4.}] The collapsed data derived from Eq.(\ref{eq.11}),
with $\nu=1.7$, $z=4.8$ and $T_g/T_c=0.84$. 100 curves are
plotted for $T>T_g$ and $T<T_g$. Inset is the original experimental
result of Ref.\cite{18}.

\end{description}
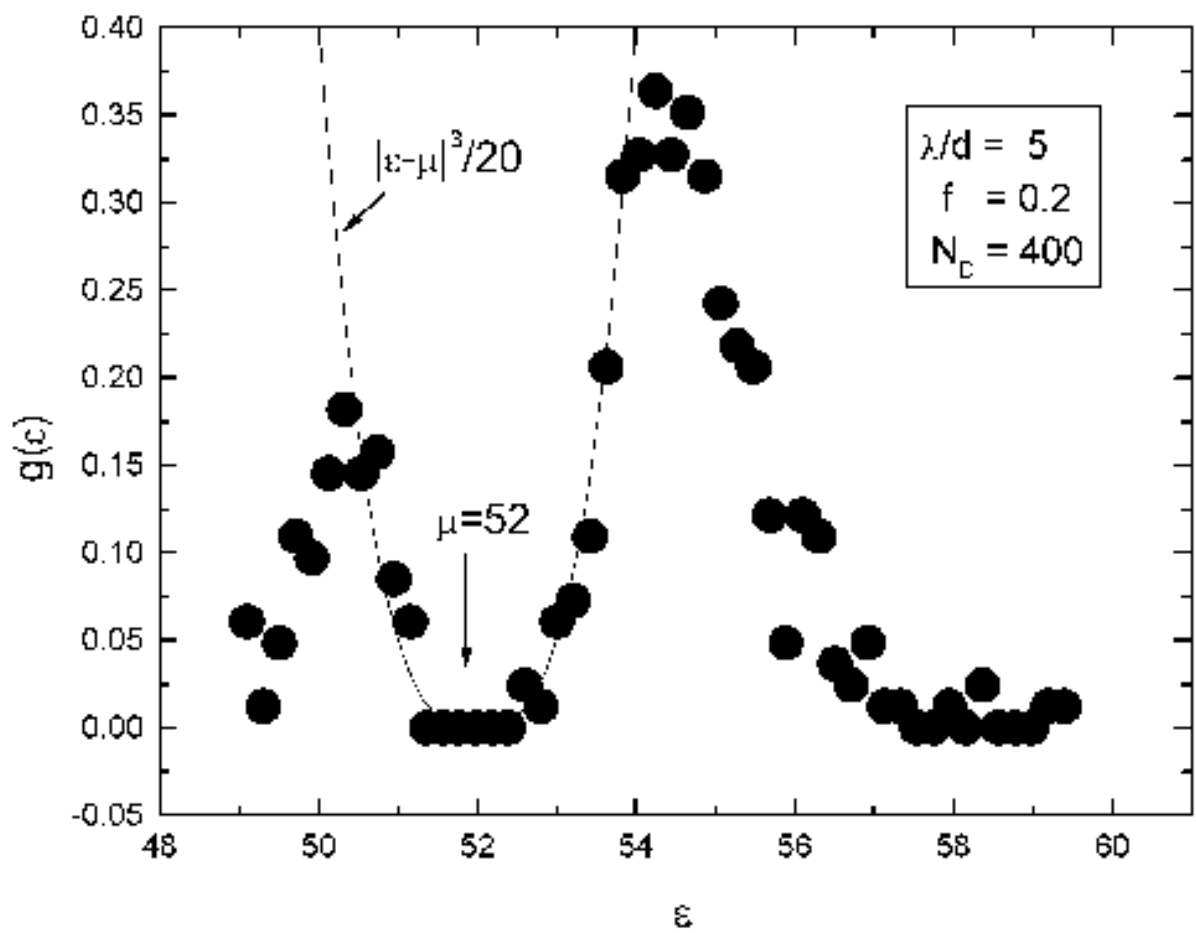
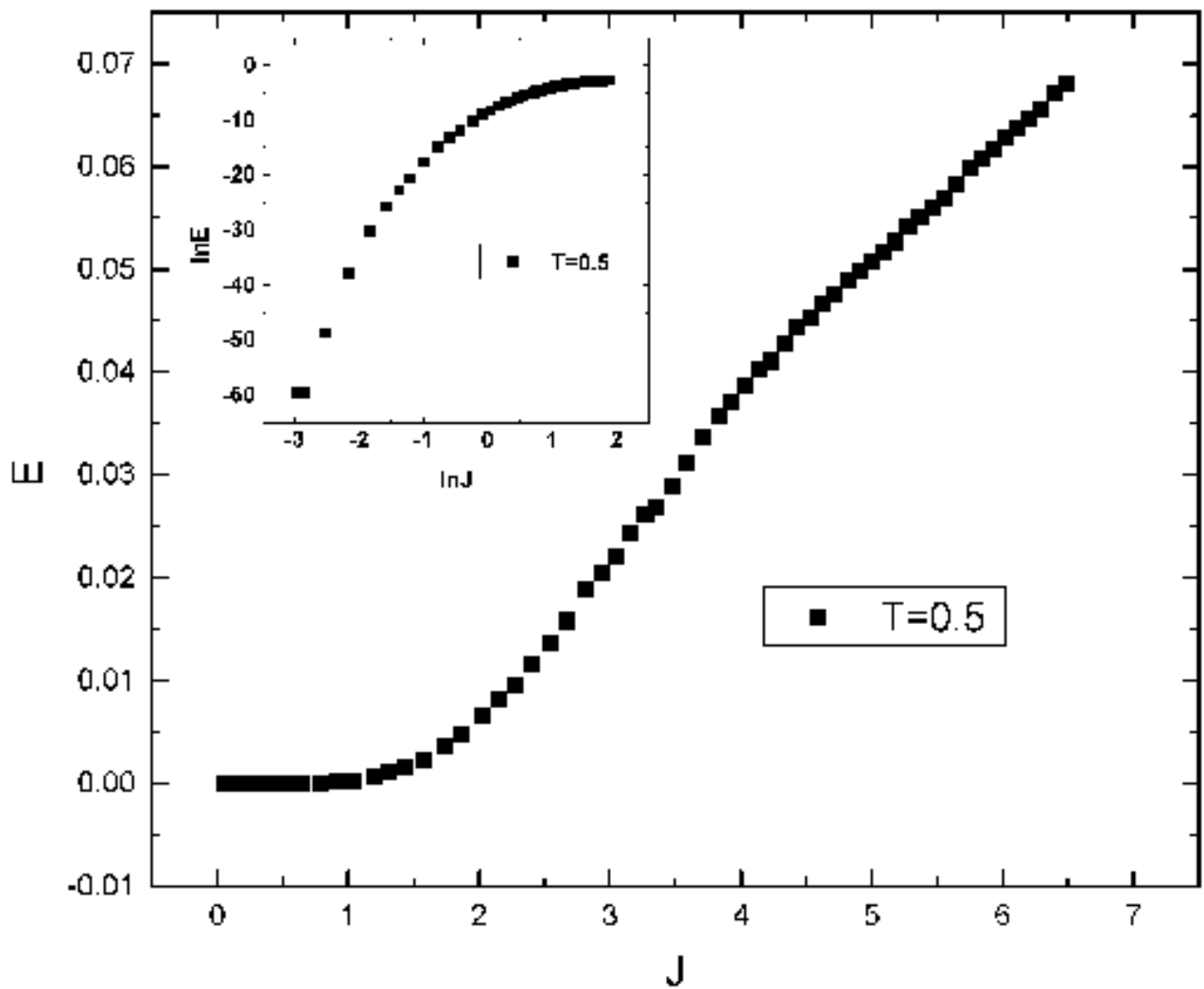
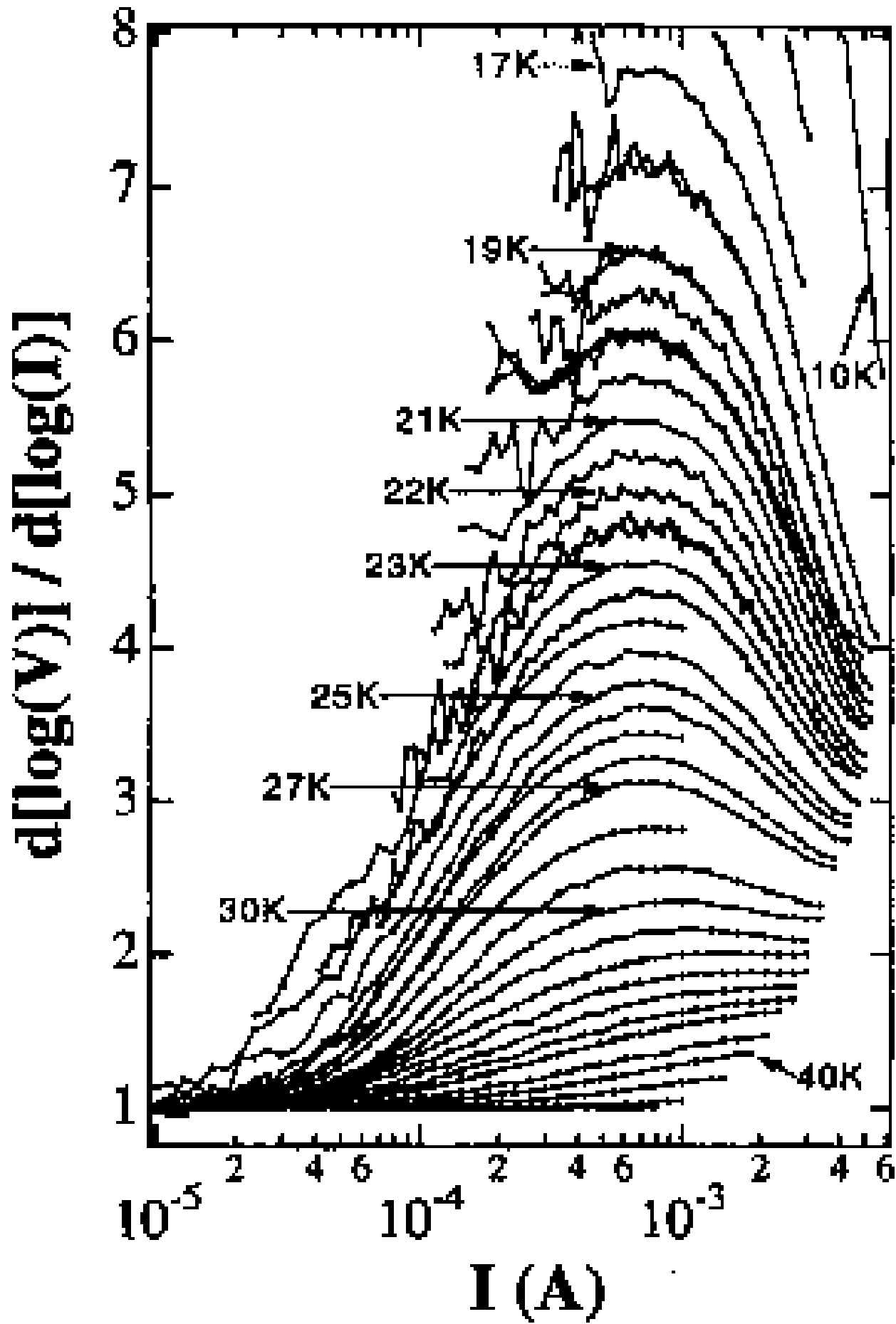
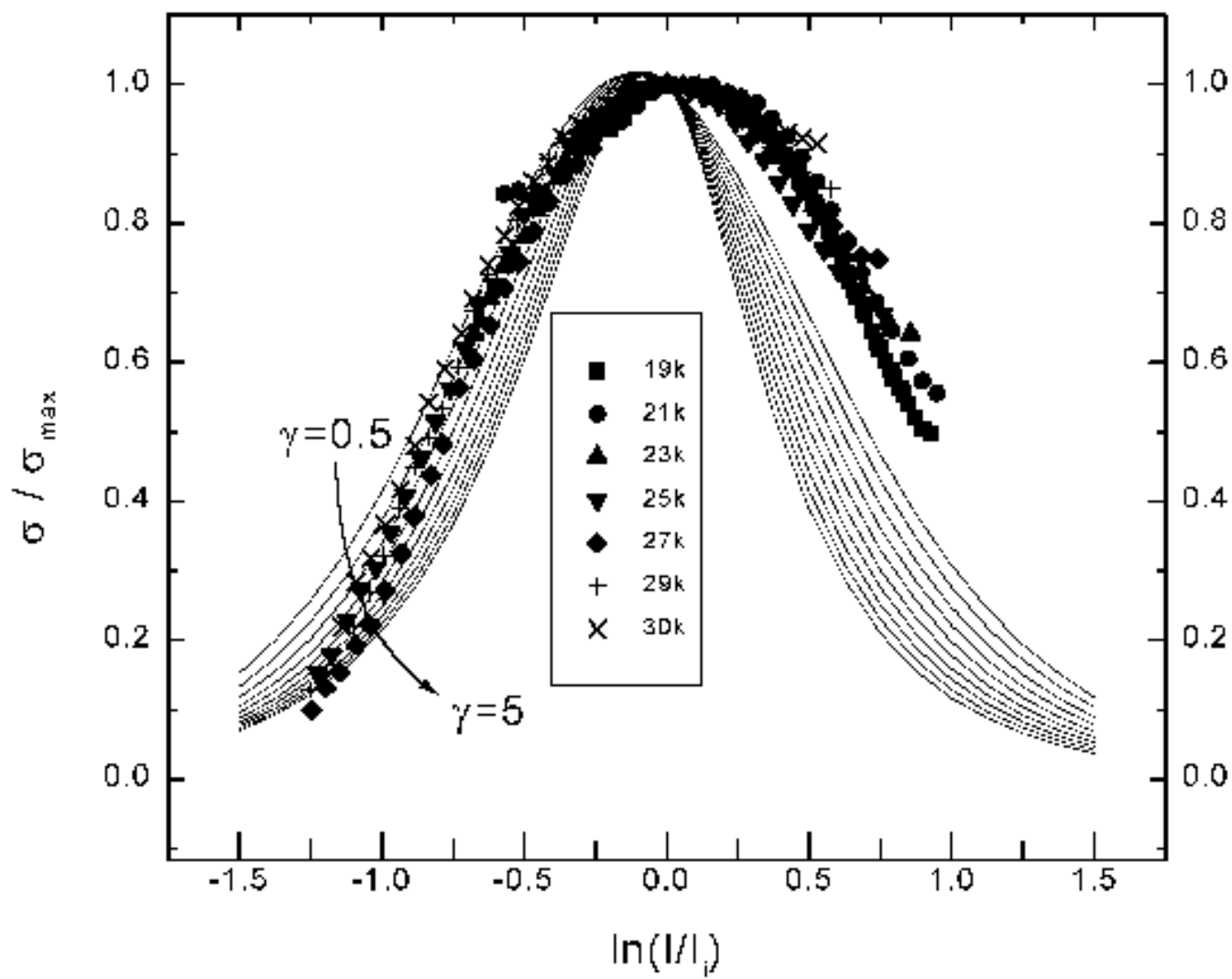
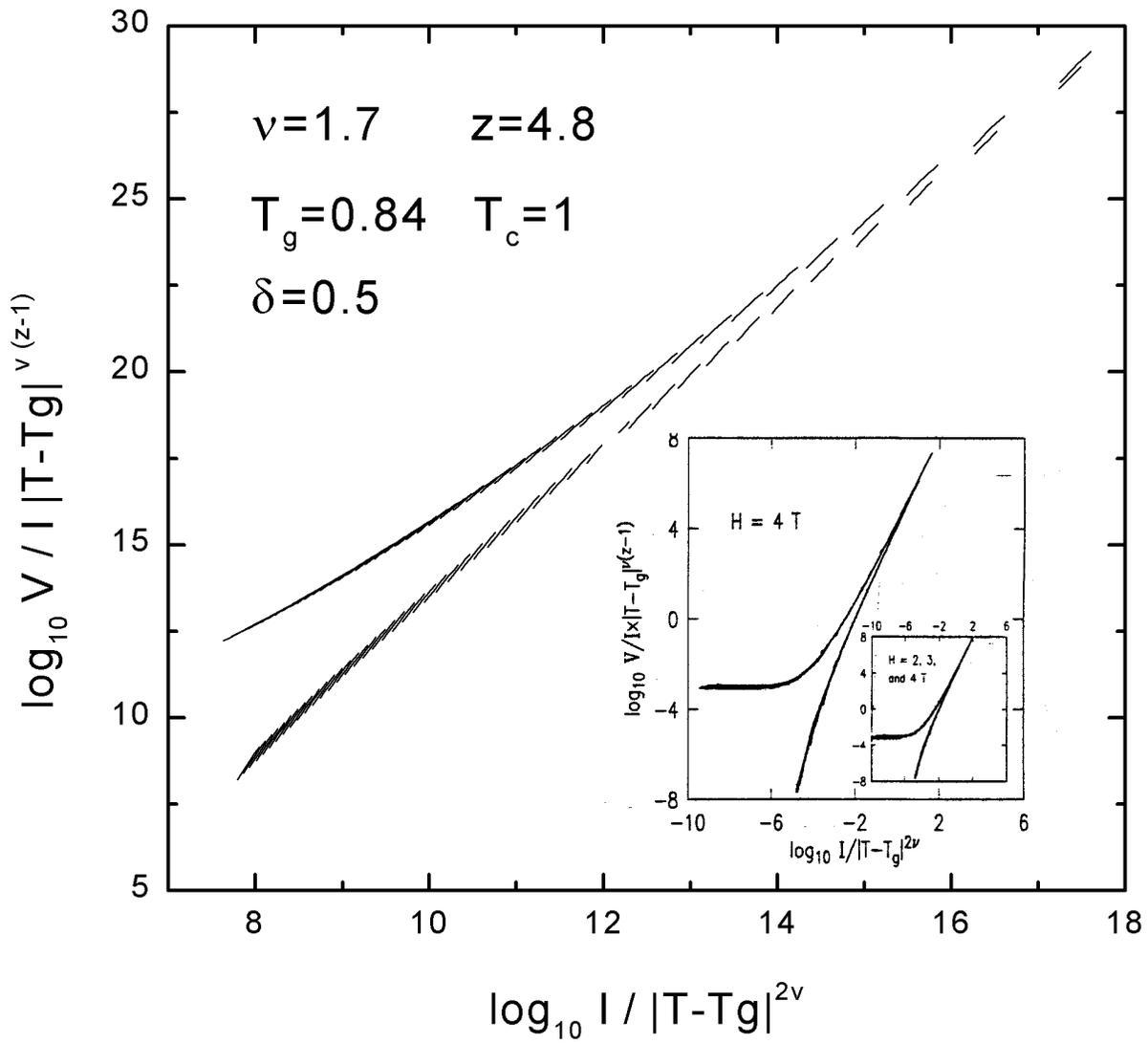
\end{document}